# Bootstrapping a Lexicon for Emotional Arousal in Software Engineering

Mika V. Mäntylä[*], Nicole Novielli[#], Filippo Lanubile[#], Maëlick Claes[*], Miikka Kuutila[*]
[*]{mika.mantyla, maelick.claes, miikka.kuutila}@oulu.fi, M3S/ITEE, University of Oulu, Finland
[#]{nicole.novielli, filippo.lanubile}@uniba.it, University of Bari, Italy

*Abstract*—Emotional arousal increases activation and performance but may also lead to burnout in software development. We present the first version of a Software Engineering Arousal lexicon (SEA) that is specifically designed to address the problem of emotional arousal in the software developer ecosystem. SEA is built using a bootstrapping approach that combines word embedding model trained on issue-tracking data and manual scoring of items in the lexicon. We show that our lexicon is able to differentiate between issue priorities, which are a source of emotional activation and then act as a proxy for arousal. The best performance is obtained by combining SEA (428 words) with a previously created general purpose lexicon by Warriner et al. (13,915 words) and it achieves Cohen's d effect sizes up to 0.5.

I. INTRODUCTION

Emotional arousal increases alertness, activation and improves software engineers' performance [1],[2]. Source of arousal in software engineering can be time pressure or task importance for example. According to Yerkes–Dodson law [3], the relationship between arousal and performance follows an inverted U-shape, with middle arousal leading to optimal performance while too low or high arousal leading to suboptimal performance. Furthermore, arousal caused by increased and prolonged pressure can also lead to burnout in software development [4]. The role of arousal in productivity and burnout in software engineering motivates our quest in developing tools for arousal detection.

This study fits within the growing research trend that recently emerged to study the role of emotion awareness in software engineering [10], by applying sentiment analysis to the content available in social coding sites [5][6], issue tracking systems [7], technical question and answering sites [8], and app reviews [11]. What the aforementioned studies have in common is that they rely on sentiment polarity (i.e. the positive, neutral or negative orientation of a text [9]) as the only dimension to operationalize affective states.

However, affect is a complex phenomenon and psychologists worked at decoding emotions for decades, developing theories aiming at the classification of emotions and their effects [12]. According to the 'circumplex model' of affect [13], emotions are distributed in a bi-dimensional representation schema, including valence on the horizontal axis (pleasant vs. unpleasant) and arousal on the vertical axis (activation vs. deactivation). Given the wide variety of affective states expressed by developers [14], valence, if employed alone, might be unreliable for mining the emotions of programmers from their technical contributions [15]. For example, detecting the negative valence of a communication trace is not enough to distinguish between anger (high arousal & negative valence) and sadness (low arousal & negative valence). Work of Murgia et al [14] also investigates dimensions beyond valence but their basis is the theory of discrete emotions by Shaver et al. [16] while we follow the dimensional theory of emotions stemming from Russel [13].

Recent research has raised concerns on the use of publicly available sentiment analysis tools for empirical software engineering [17],[18], which have been trained on non-software engineering documents, such as movie reviews or posts crawled from general-purpose social media (e.g., Twitter and YouTube). In particular, Jongeling et al. [18] assessed the predictions of popular sentiment analysis tools showing that not only these tools do not agree with human annotation of developers' communication channels, but they also disagree between themselves. As such, adapting existing sentiment analysis tools and lexicons become crucial to provide valid conclusions in software engineering studies.

In this paper, we aim at advancing the state of the art on emotion mining in software engineering by addressing these issues. We describe the bootstrapping of a software engineering specific lexicon for arousal, which we release for research purposes[1]. We validate our approach by testing our lexicon's ability to differentiate issues with different priorities, which should create different levels of arousal.

II. RESEARCH METHODS

Sentiment analysis is the study of the valence (also named as polarity) of a text [9]. Publicly available sentiment analysis tools, such as SentiStrength [19], rely on sentiment lexicons, that is, large collections of words with information about association to sentiment (i.e., prior association). The overall sentiment of a text, its valence, is therefore computed based on such prior associations of the contained words.

Sentiment lexicons can be manually annotated, either by experts or by the crowd [20]. Alternatively, semi- or fully-automated approaches have been adopted to estimate the sentiment associations of a large collection of words or phrases, starting from a small initial set of manually annotated data [21]. Among semi-automated techniques, ongoing research is currently addressing approaches that leverage distributional models of lexical semantics, also known as *word embedding* [22], to bootstrap lexicons for a particular domain or language [23][24]. In particular, Passaro et al used bootstrapping to create an emotional lexicon for Italian [24]. Similarly, we adopt a bootstrapping approach by jointly leveraging word embedding and human annotation, which consists of the following steps.

---

[1] Software Engineering Arousal (SEA) Lexicon can be downloaded from Figshare at https://doi.org/10.6084/m9.figshare.4781188.v1

First, we studied a general-purpose lexicon by Warriner et al. [20], containing arousal scores for roughly 14,000 English words, to select seed words potentially denoting high or low arousal in software engineering context. Words were also checked for the number of occurrences in the data set by Ortu et al. [25], containing 700,000 issue reports from Apache Jira issue tracking system. For the first 10 high and low arousal words, we required more than 100 occurrences and for the next 10 high and low words we required more than 1,000 occurrences. These limits guarantee that the selected seeds occur in software engineering data.

Second, we included additional seed words potentially indicative of arousal from different sources: 1) surveys asking about general job demands [26],[27], 2) surveys about specific tasks, i.e. NASA TLX [28], 3) Russell's circumplex model of affect [13] extended in [29], 4) words from a text analysis application called Linguistic Inquiry and Word Count [30] about anxiety, time, and achievement, 5) profanities used in software engineering [31], 6) brainstormed words that we thought should be included.

Third, we searched and included relevant synonyms for the seed words from WordNet [32], a semantically structured lexical database for English that is very popular among researchers in computational linguistics and text analysis.

Fourth, we searched for further synonyms by studying words that were used similarly as our seeds in the data set by Ortu et al [25], i.e. words used in similar contexts. To this aim, we exploited the GloVe algorithm [33], implemented in R [34] as part of the text mining package text2vec [35]. We trained our word embedding vector space with 300 dimension vectors, using the skip-gram model implemented by the tool, and choosing a window size of 10 words. The word pair technique, which corresponds to a word window size of two, has been previously proposed to find semantically similar words from the source code and its comments [36]. For each seed, we investigated its 10 closest neighbors based on their cosine similarity. This allowed us to find and include domain specific words indicative of high and low arousal, for example word "soon" was used similarly as word "asap". At this stage, we also excluded seed words as we learned that many words that we identified as potentially related to performing something fast (indicative of high emotional arousal for humans) were instead used to describe the performance of software rather than software engineers. In the end, 350 words were included by synonym searching from Ortu data set or via WordNet while 78 came from seed sources.

Fifth, the words collected in the first four steps (n=428), were evaluated by two authors who had not been involved in the previous steps. Figure 1 shows an excerpt from the word rating spreadsheet We modified the task description from Warriner et al [20] to suit our needs, as follows:

"*You are invited to take part in the study that is investigating how software developers feel when they use different types of words in Jira issue reports or issue report comments. You will use a scale to rate how you think a software developer felt when using each word. There will be approximately 450 words. The scale ranges from 1 calm to 9 excited. At one extreme of this scale, software developers feel relaxed, calm, sluggish, dull, sleepy, or unaroused (rating 1). The other end of the scale is when they are stimulated, excited, frenzied, jittery, wide-awake, or aroused (rating 9). The numbers also allow you to describe intermediate feelings of calmness/arousal, by selecting any of the other feelings. If you think they feel completely neutral, not excited nor at all calm, select the middle of the scale (rating 5).*

*For each evaluated word, we have included a list of ten words which are used in a similar way in the issue reports and comments. The purpose of these ten words is to help you rate each word more accurately as they offer clues how each word is used. Please note that the number behind each of the ten words measures the similarity in the context which the word appears in (1.00 being perfect similarity and 0 none existent).*

*Please work at a rapid pace and don't spend too much time thinking about each word.*"

Finally, we evaluated our lexicon by comparing its ability to differentiate between different defect severities, also with respect to the results obtained by Mantyla et al. [37].

| Word | 1 (calm) - 9 (excited) | Frequency | Words used in similar way | |
|---|---|---|---|---|
| scary | | 632 | messy - 0.54 | annoying - 0.52 |
| seeing | | 15103 | saw - 0.70 | happening - 0.69 |
| seem | | 32112 | don - 0.80 | seems - 0.80 |
| seemed | | 4931 | seems - 0.67 | seem - 0.64 |
| seems | | 124803 | but - 0.90 | though - 0.88 |
| sense | | 34062 | makes - 0.85 | think - 0.83 |
| sensible | | 1779 | reasonable - 0.62 | defaults - 0.53 |
| september | | 796 | october - 0.81 | august - 0.76 |
| serious | | 4051 | problems - 0.54 | significant - 0.53 |
| severe | | 4235 | solrexception - 0.39 | grave - 0.39 |
| severely | | 274 | impacted - 0.47 | seriously - 0.41 |
| shit | | 225 | crazy - 0.37 | spew - 0.32 |
| shortly | | 5894 | soon - 0.75 | tomorrow - 0.62 |
| showstopper | | 366 | blocker - 0.46 | serious - 0.38 |
| sick | | 174 | vacation - 0.36 | traveling - 0.35 |
| simple | | 67481 | example - 0.66 | easy - 0.61 |
| situation | | 13828 | scenario - 0.72 | situations - 0.71 |
| situations | | 4988 | scenarios - 0.71 | situation - 0.71 |
| skeptical | | 177 | worried - 0.47 | wary - 0.45 |
| small | | 28099 | large - 0.71 | tiny - 0.63 |
| solution | | 56005 | way - 0.77 | possible - 0.76 |
| solvable | | 179 | fixable - 0.42 | remedied - 0.41 |
| someday | | 364 | anytime - 0.40 | hoping - 0.35 |
| someone | | 27016 | somebody - 0.84 | anyone - 0.78 |
| sometime | | 209 | week - 0.64 | soon - 0.62 |
| soon | | 21994 | shortly - 0.75 | ll - 0.74 |

**Figure 1 Snapshot of the word rating spreadsheet**

## III. RESULTS

### A. Ratings

TABLE I. shows statistics about the rating process. The two human raters produced very similar mean scores (5.42 and 5.35) and standard deviations (1.53 and 1.32). The correlation (Pearson) between ratings is highly significant with alpha level 0.001 (p-value 1.496e-11). The correlation coefficient is not very high at 0.32. Warriner does not report correlation between individuals but they report the Pearson correlations between several group averages and they fluctuate between 0.47 and 0.52 for arousal, while being much higher for Valence 0.79 and 0.83. Murgia et al. [14] performed a similar rating process and reported, as interrater agreement, Kappa values of 0.51, 0.19, -0.01, 0.06, 0.18, 0.10, for emotions love, joy, surprise, anger, sadness, and fear respectively. The Kappa value for our ratings is 0.32, which is higher than what was obtained for 5 out of 6 emotions studied by Murgia et al. However, we used 1-9 scale to rate the existence of emotional arousal for individual words while Murgia et al. used a dichotomous scale to rate each emotion in a piece of text consisting of one or more sentences. Regardless of these differences between the studies, Kappa values are comparable.

TABLE I. STATISTICS ABOUT RATINGS

| Mean | 5.42 (R1), 5.35 (R1) |
|---|---|
| Std deviation | 1.53 (R1), 1.32 (R2) |
| Correlation (Pearson) | 0.32 (p-value 1.496e-11) |
| Kappa (Weighted) | 0.32 |
| Perfect agreement (1-9 scale) | 28% |
| Perfect agreement or off-by-one (1-9 scale) | 69% |
| Opposite ratings (e.g. R1>5 and R2<5) | 14% |

Raters reported having difficulties rating words with multiple possible contexts. One of them also reported frequently giving neutral scores, and rarely giving extreme score values. The same rater also reported having given slightly different scores to similar words with different endings, such as 'annoyed' and 'annoying'. This is reasonable since different inflected forms of the same lemma may convey different information about sentiment. It is the case, for example, of the SentiStrength lexicon where inflected forms of the same words hold different sentiment polarity and strength [19].

Finally, both raters reported about comparing the word they were rating to the few previously rated words. Thus, the alphabetical order in which words were listed might influence the results.

### B. Validation

We validated our approach by comparing the lexicon's ability to differentiate between issue priorities in the data set by Ortu et al. Based on psychological literature, urgency [38] and increases in rewards [39] or punishments [3] increase emotional arousal. In this work, we make the assumption that higher priority issues are more urgent and fixing them results in higher reward in terms of system quality improvement. Thus, we assume that working on higher priority issues is associated with elevated emotional arousal. As such, if our lexicon is useful in practice, it should be able to differentiate between issue priorities.

We compare and replicate our approach with the one presented by Mantyla et al. [37] where the authors used the general-purpose lexicon by Warriner et al. Consistently with their approach, we measure the arousal score of a given text unit as a function of its words' individual arousal scores. In particular, a text overall arousal score is computed by considering the two words with the *max* and *min* arousal. If *max* is lower than the average value or *min* is higher than the average value then we set the *max* or *min* to the average of all words of the lexicon. If the lexicon does not match with any elements of the text unit, the score is given no value and the text unit is not considered in the statistical analysis. We compute the arousal scores (*max+min*) for each issue element, namely its Title, Description, All comments, First comment and Last Comment. Hence, we compare the arousal score across the five issue priorities supported by Jira. To this aim, we perform t-tests to assess the statistical significance of differences and compute the Cohen's d to assess the effect size.

TABLE II. shows Cohen's d effect size between issue priorities and TABLE III. shows the alpha levels of the t-test. The tables present measures about different parts of issue reports, as indicated in the first column. The second columns indicate the lexicons from which results have been obtained: [37] indicates the original results of the study which we have replicated, where the original lexicon by Warriner et al. was used; SEA indicates the lexicon developed in this paper; united approach ([37] +SEA) considers the adjusted lexicon obtained by summing up the scores produced by Mantyla et al. [37] and SEA as follows: [37] + (SEA_Score – SEA_AVG). This modifies the original score of [37] by increasing or decreasing it depending on the SEA score. We subtract the global SEA average to allow easier backward comparison to [37]. Bold numbers indicate which of three approaches provides the highest Cohen's d.

TABLE II. shows that in the comparison of polar opposites Blocker vs. Trivial united approach is always the best and produces Cohen's d values up to medium effect size (0.5). If we look only to adjacent defect priorities, i.e. the four rightmost columns in TABLE II. , and interpret the results as a pair-wise competition, we can see that the united approach ([37] + SEA) beats the original approach of [37] with 15 victories vs 5. This demonstrates clear benefits of combining general-purpose and software engineering specific lexicons. Competition between SEA and original [37] shows that original is slightly better with 11 victories vs 9. More detailed comparison between SEA and original shows that SEA scores 8 victories with higher defect severities, Blocker-Critical-Major while original has only 2 victories. With lower defect severities, Major-Minor-Trivial the roles are reversed and original wins SEA with 9 vs 1 victories. Thus, our lexicon makes a partly strong contribution to differentiating the very important defects from the others. At the same time the original approach is partially strong in differentiating between minor and trivial defects.

The original lexicon used in [37] contains 13,915 words while our lexicon had only 428 words. Thus, with a much smaller domain specific lexicon we demonstrate similar performance as the original, and when we combine the approaches we demonstrate superior performance.

TABLE II. COHEN'S D BETWEEN ISSUE PRIORITIES (Bold indicates which is the best: [37], SEA, or [37]+SEA)

| Field | Lexicon | Blocker-Trivial | Blocker-Critical | Critical-Major | Major-Minor | Minor-Trivial |
|---|---|---|---|---|---|---|
| Title | [37] | 0.2411 | **0.0134** | 0.0751 | 0.0696 | **0.0784** |
| | SEA | 0.1267 | 0.0106 | 0.0718 | 0.0323 | 0.0169 |
| | [37]+SEA | **0.3643** | -0.0032 | **0.2212** | **0.0884** | 0.0686 |
| Desc | [37] | 0.3240 | -0.0099 | 0.1173 | **0.0482** | 0.1609 |
| | SEA | 0.2875 | **0.0649** | 0.1489 | 0.0149 | 0.0798 |
| | [37]+SEA | **0.3954** | 0.0324 | **0.1878** | 0.0371 | **0.1731** |
| All comments | [37] | 0.3541 | 0.0041 | 0.1233 | 0.0332 | **0.1843** |
| | SEA | 0.3880 | 0.0106 | 0.2069 | 0.0417 | 0.0755 |
| | [37]+SEA | **0.5070** | **0.0835** | **0.2286** | **0.0583** | 0.1797 |
| First comment | [37] | 0.2753 | -0.0176 | 0.0833 | 0.0565 | 0.1528 |
| | SEA | 0.3438 | **0.1065** | 0.1558 | 0.0380 | 0.0813 |
| | [37]+SEA | **0.4209** | 0.0595 | **0.1647** | **0.0624** | **0.1754** |
| Last comment | [37] | 0.1804 | -0.0018 | 0.0514 | 0.0223 | **0.1086** |
| | SEA | 0.2337 | **0.0412** | 0.1168 | 0.0457 | 0.0423 |
| | [37]+SEA | **0.2763** | 0.0049 | **0.1386** | **0.0461** | 0.1018 |

TABLE III. T-TEST P-VALUES BETWEEN ISSUE PRIORITIES (BOLD < 0.001, *ITALIC* <0.01, UNDERLINE <0.05)

| Field | Lexicon | Blocker-Trivial | Blocker-Critical | Critical-Major | Major-Minor | Minor-Trivial |
|---|---|---|---|---|---|---|
| Title | [37] | **< 2.2e-16** | 0.1364 | **< 2.2e-16** | **< 2.2e-16** | **< 2.2e-16** |
| | SEA | **0.00012** | 0.729 | *0.00303* | **0.00063** | 0.4509 |
| | [37]+SEA | **< 2.2e-16** | 0.9174 | **< 2.2e-16** | **< 2.2e-16** | *0.0012* |
| Desc | [37] | **< 2.2e-16** | 0.2626 | **< 2.2e-16** | **< 2.2e-16** | **< 2.2e-16** |
| | SEA | **< 2.2e-16** | **3.091e-06** | **< 2.2e-16** | **0.00096** | **1.27e-11** |
| | [37]+SEA | **< 2.2e-16** | **<2.2e-16** | **< 2.2e-16** | **< 2.2e-16** | **< 2.2e-16** |
| All comments | [37] | **< 2.2e-16** | 0.6688 | **< 2.2e-16** | **< 2.2e-16** | **< 2.2e-16** |
| | SEA | **< 2.2e-16** | **< 2.2e-16** | **< 2.2e-16** | **< 2.2e-16** | **8.986e-10** |
| | [37]+SEA | **< 2.2e-16** | **7.373e-12** | **< 2.2e-16** | **< 2.2e-16** | **< 2.2e-16** |
| First comment | [37] | **< 2.2e-16** | 0.0756 | **< 2.2e-16** | **< 2.2e-16** | **< 2.2e-16** |
| | SEA | **< 2.2e-16** | **2.121e-10** | **< 2.2e-16** | **7.951e-10** | **2.583e-07** |
| | [37]+SEA | **< 2.2e-16** | **0.00037** | **< 2.2e-16** | **< 2.2e-16** | **< 2.2e-16** |
| Last comment | [37] | **< 2.2e-16** | 0.8576 | **3.608e-13** | **3.145e-09** | **< 2.2e-16** |
| | SEA | **< 2.2e-16** | <u>0.04448</u> | **2.689e-16** | **1.331e-10** | <u>0.01864</u> |
| | [37]+SEA | **< 2.2e-16** | 0.8115 | **< 2.2e-16** | **1.112e-10** | **6.878e-08** |

## IV. CONCLUSION AND FUTURE WORK

We presented a Software Engineering Arousal lexicon (SEA) specifically designed to address the problem of detecting emotional arousal in the software developer ecosystem. To the best of our knowledge, SEA is the first publicly available arousal lexicon developed for supporting research on emotional awareness in software engineering. SEA is built using a bootstrapping approach that combines word embedding model trained on issue-tracking data and manual scoring of items in the lexicon.

We evaluated SEA based on its ability to distinguish between issues based on their priority, which is a source of emotional activation and, therefore, acts as a proxy for arousal. The unified approach combining SEA (428 words) with a general purpose lexicon by Warriner et al. (13,915 words) offers clear improvement over previous work. Although 259 SEA words also appear in Warriner, the arousal scores in SEA are changed through the annotation performed in the present study. This means that a sentiment lexicon tuned for software development may improve emotion mining in empirical software engineering studies.

In future work, we plan to replicate the bootstrapping approach to fine-tune scores for other emotion dimensions that have been previously addressed in software engineering research, namely valence and dominance, for which scores are provided in the lexicon by Warriner et al.

We acknowledge some limitations of the current study. First of all, we computed the arousal score of each text unit in our dataset by only considering prior association of words with their individual arousal scores in the lexicon. We are aware of the need for dealing with negations, degree adverbs, intensifier, and modals, which act as modifiers on both the word meaning and associated sentiment. While a general consensus is observed in literature about the effect of modifiers on valence (e.g. a negation usually reverses the positive/negative orientation of words in its scope) [40], less attention has been devoted to the impact of modifiers on emotional arousal. In our future work, we plan to refine our approach to deal with textual modifiers.

We intend to use the lexicon to study the impact of software developers' arousal in the context of a time pressure environment. To achieve this goal, we will combine multiple metrics, such as changes in timestamp and periods of activity, with arousal. Relating these metrics with deadlines, in different time pressure environments, would give insights on how the policies used by different software projects impact developer productivity and health.

Finally, in our future research we plan to address the problem of correctly identifying the emotion target, that is the interlocutor (i.e., another developer), an object (i.e., a tool, a language, a software), or the writer itself, in line with ongoing research on affect detection in software engineering [41] and with aspect oriented sentiment analysis [42]. Being able to clearly identify the emotion target is crucial for differentiating between texts talking about actual high-arousal triggering situations (as in '*we need to be fast in addressing this issue*', directed to *other* developers) and neutral scenarios where people simply describe the properties of an *object* (as in '*a fast software system*').

ACKNOWLEDGMENT

Authors at the University of Oulu have been supported by the Academy of Finland Grant no 298020. Authors at the University of Bari have been supported by the project EmoQuest, funded by MIUR under the SIR program.